\documentstyle[amsmath,epsf,epsfig,preprint,aps,amssymb]{revtex}
\draft \preprint{SNUTP 03-015}
\begin{document}
\title{\Large\bf Higgsino mass matrix ansatz for MSSM}
\author{
Kang-Sin Choi$^{(a,b)}$\footnote{ugha@phya.snu.ac.kr}, 
Ki-Young Choi$^{(a,b)}$\footnote{ckysky@phya.snu.ac.kr},
Kyuwan Hwang$^{(a)}$\footnote{kwhwang@phya.snu.ac.kr},
and Jihn E.  Kim$^{(a,b)}$\footnote{jekim@phyp.snu.ac.kr}
}
\address{
$^{(a)}$School of Physics, Seoul National University, Seoul
151-747,
Korea\\
$^{(b)}$Physikalisches Institut,
Universit\"at Bonn, Nussallee 12, D53115, Bonn, Germany}
\maketitle

\begin{abstract}
An ansatz, Det.~$M_{\tilde H}=0,$ for the Higgsino mass 
matrix in string orbifold trinification is suggested 
toward the minimal supersymmetric standard model(MSSM). 
Small instanton solutions effective around the GUT scale 
can fulfil this condition. An argument that 
the couplings contain a moduli field is given for a 
dynamical realization of this Higgsino mass matrix ansatz.
\\
\vskip 0.5cm\noindent [Key words: MSSM, orbifold,
trinification, Higgsino mass]
\end{abstract}

\pacs{11.30.Hv, 11.25.Mj, 11.30.Ly}

\newpage
\newcommand{\bea}{\begin{eqnarray}}
\newcommand{\eea}{\end{eqnarray}}
\def\beq{\begin{equation}}
\def\eeq{\end{equation}}

\def\one{\bf 1}
\def\two{\bf 2}
\def\five{\bf 5}
\def\ten{\bf 10}
\def\tenb{\overline{\bf 10}}
\def\fiveb{\overline{\bf 5}}
\def\threeb{{\bf\overline{3}}}
\def\three{{\bf 3}}
\def\fb{{\overline{F}\,}}
\def\hb{{\overline{h}}}
\def\Hb{{\overline{H}\,}}

\def\To{${\cal T}_1$\ }
\def\Tt{${\cal T}_2$\ }
\def\Z{${\cal Z}$\ }
\def\ot{\otimes}
\def\tr{{\rm tr}}
\def\sinw{{\sin^2 \theta_W}}

\def\slash#1{#1\!\!\!\!\!\!/}

\newcommand{\dis}[1]{\begin{equation}\begin{split}#1\end{split}\end{equation}}
\newcommand{\beqa}[1]{\begin{eqnarray}#1\end{eqnarray}}

\def\be{\begin{equation}}
\def\ee{\end{equation}}
\def\ben{\begin{enumerate}}
\def\een{\end{enumerate}}
\def\lsl{ l \hspace{-0.45 em}/}
\def\ksl{ k \hspace{-0.45 em}/}
\def\qsl{ q \hspace{-0.45 em}/}
\def\psl{ p \hspace{-0.45 em}/}
\def\ppsl{ p' \hspace{-0.70 em}/}
\def\dsl{ \partial \hspace{-0.45 em}/}
\def\Dsl{ D \hspace{-0.55 em}/}
\def\matrix{ \left(\begin{array} \end{array} \right) }

\def\ma{m_A}
\def\mf{m_f}
\def\mz{m_Z}
\def\mw{m_W}
\def\ml{m_l}
\def\ms{m_S}
\def\dag{\dagger}

 \def\NCA{{\em Nuovo Cimento} }
 \def\NIM{{\em Nucl. Instrum. Methods} }
 \def\NIMA{{\em Nucl. Instrum. Methods} A }
 \def\NP{{\em Nucl. Phys.} }
 \def\NPB{{\em Nucl. Phys.} B }
 \def\PL{{\em Phys. Lett.} }
 \def\PLB{{\em Phys. Lett.} B }
 \def\PRL{{\em Phys. Rev. Lett.} }
 \def\PRD{{\em Phys. Rev.} D }
 \def\PR{{\em Phys. Rev.} }
 \def\RMP{{\em Rev. Mod. Phys.} }
 \def\ZPC{{\em Z. Phys.} C }
 \def\PHYSICA{{\em Physica} D }
 \def\CMP{{\em Commun. Math. Phys.} }
 \def\PREP{{\em Phys. Rep.} }
 \def\JMP{{\em J. Math. Phys.} }
 \def\CQG{{\em Class. Quant. Grav.} }
 \def\ANN{{\em Annals of Physics} }
 \def\APP{{\em Acta Phys. Polon.} }
 \def\RPP{{\em Rep. Prog. Phys.} }

\section{Introduction}

The current studies on supersymmetric standard model are based on
an implicit assumption that the minimal supersymmetric standard
model(MSSM) may be derivable from a fundamental theory valid at a
very high energy scale. We can
list three important theoretical problems in such low energy
supergravity models:

\begin{itemize}
\item The $\mu$ problem~\cite{kn84},

\item The doublet-triplet splitting problem, and

\item The problem of the MSSM construction from a fundamental theory.

\end{itemize}

The twenty year old $\mu$ problem has two well-known solutions, a
dimension-4 superpotential~\cite{kn84} along with a Peccei-Quinn
symmetry and a softly broken supergravity effect supplied with
some symmetry~\cite{giudice}. In any case, there must be some kind
of symmetry to forbid a large $\mu$ term~\cite{kn94}. Ever since
writing a supersymmetric GUT~\cite{susygut}, the doublet-triplet
splitting problem has become one of the most fundamental problems
in supersymmetric model building. In string orbifolds, there are
examples that colored chiral fields beyond those in the standard
model(SM) do not appear~\cite{iknq}, which is a kind of a solution
for the doublet-triplet splitting problem. However, the old {\it
standard-like} models suffered from the unification problem in
that generally the bare value of $\sin^2\theta_W\ne
\frac38$~\cite{kim03}. In addition, the number of Higgs doublets
is more than one pair of $H_u$ and $H_d$. This problem of more
than one pair of Higgs doublets violates the gauge coupling
unification. Only the MSSM spectrum, i.e. only one pair of Higgs
doublets at the electroweak scale, is consistent with the gauge
coupling unification with the bare value of 
$\sin^2\theta_W^0=\frac38$. 

Therefore, this problem is listed above as
the problem of {\it MSSM construction}. Certainly this MSSM
problem is a kind of the $\mu$ problem, since we must have a light
Higgs doublet pair. So, the three problems
listed above are all related.

In string constructions, usually there appear many Higgs doublets.
In particular, the most attractive $Z_3$ orbifold models allow
multiples of three pairs of Higgs doublets. Recently, it has been
pointed out that $Z_3$ orbifold models with a Wilson line(s)
resulting in three families of trinification spectrum in the gauge
group $SU(3)_1\times SU(3)_2\times SU(3)_3$,
\begin{equation}\label{trinification}
\Psi_{\rm tri}\equiv
{\bf (\bar 3,3,1)+(1,\bar 3,3)+(3,1,\bar 3)}
\end{equation}
can be the connecting theory between string scale and the MSSM
scale~\cite{kim03}. Without a Wilson line, it was not possible to
obtain the trinification spectrum~\cite{tables}.

To discuss the fields in terms of the SM gauge quantum numbers,
let us follow the notation given in~\cite{ck03,su38},
\begin{eqnarray}
{\bf (\bar 3,3,1)}=\Psi_l\longrightarrow
  \Psi_{(\bar M,I,0)}&=& \Psi_{(\bar 1,i,0)}(H_1)_{-\frac12}+
  \Psi_{(\bar 2,i,0)}(H_2)_{+\frac12}
  + \Psi_{(\bar 3,i,0)}(l)_{-\frac12}\nonumber\\
  &+&\Psi_{(\bar 1,3,0)}(N_5)_0+
  \Psi_{(\bar 2,3,0)}(e^+)_{+1}+ \Psi_{(\bar 3,3,0)}(N_{10})_0
\label{rep1}
\\
{\bf (1,\bar 3,3)}=\Psi_q\longrightarrow
  \Psi_{(0,\bar I,\alpha)}\ &=& \Psi_{(0, \bar i,\alpha)}
  (q)_{+\frac16}+ \Psi_{(0,\bar 3,\alpha)}(D)_{-\frac13}
\label{rep2}
\\
{\bf (3,1,\bar 3)}=\Psi_a\longrightarrow
  \Psi_{(M,0,\bar\alpha)}&=& \Psi_{(1,0,\bar\alpha)}
  (d^c)_{\frac13}+ \Psi_{(2,0,\bar\alpha)}(u^c)_{-\frac23}
  + \Psi_{(3,0,\bar\alpha)}(\overline{D})_{+\frac13}
\label{rep3}
\end{eqnarray}
where $\langle N_{10}\rangle$ and $\langle N_5\rangle$
are needed to break $SU(3)^3$ down to the SM gauge group.
We introduced the above names in view of the SM fields. The above
three different representations are named carrying different {\it
humors}: {\it lepton--, quark--,} and {\it antiquark--humors}. In
these three sets of trinification spectrum, there are three pairs
of Higgs doublets.

In Ref.~\cite{ck03}, the trinification spectrum arises in the
untwisted sector. However, that model has two phenomenological
problems. One is that without $\bar\Psi_{\rm tri}$ in the spectrum
it is difficult to achieve a reasonable neutrino mass spectrum.
Also, the D-flat direction is difficult to obtain. Therefore, it
is better to have six $\Psi_{\rm tri}$'s and three $\bar\Psi_{\rm
tri}$'s instead of just three $\Psi_{\rm tri}$'s so that
$\bar\Psi_{\rm tri}$'s also participate in the gauge symmetry
breaking. In addition, at least three $\Psi_{\rm tri}$'s must come
from the twisted sector so that the mass matrix is not
antisymmetric~\cite{nilles90}. If the whole sets of $\Psi_{\rm
tri}$ and $\bar\Psi_{\rm tri}$ are added in addition to the three
$\Psi_{\rm tri}$'s for the three light families, the GUT scale
value of $\sin^2\theta_W$ is the desired $\frac38$. However, it
may be possible to obtain a realistic $\sin^2\theta_W$ even though
we start with a much smaller GUT scale value of
$\sin^2\theta_W^0$. Sometimes, it is called $\lq$optical
unification'~\cite{giedt}. In this case, one has to introduce another
parameter so that the evolution of gauge couplings change
appropriately between a scale $M_1$ and $M_{GUT}$ due to the
assumption how the vectorlike particles are removed in this
region. Certainly, this proposal is not as attractive as the
$\sin^2\theta_W^0=\frac38$ model. However, it can have its own
virtue in that it may have smaller number of particles, sacrifying
$\sin^2\theta_W^0=\frac38$. In the following, we present such a
model, present an ansatz for the Higgsino mass matrix, and
discuss some related issues.

\section{Just three more lepton humors}

In $Z_3$ orbifold compactifications, one needs
Wilson lines to have the trinification spectrum~\cite{kim03}.
One interesting model has been studied before~\cite{ck03},
but there the trinification families appear from the
untwisted sector. In general, the mass matrix of the
untwisted sector is antisymmetric from the $H$-momentum 
rule~\cite{hv87,fiqs},
and we have an unacceptable relation
$m_t=m_c$~\cite{nilles90}. Therefore, it is better to have three
families from the twisted sectors. With two Wilson lines,
indeed there exist three family models from the twisted
sector. The shift vector and Wilson lines are\footnote{We
show this just for the sake of possible realization of our 
ansatz, without worrying much about phenomenological
obstacles such as $\sin^2\theta_W^0\ne \frac38$, etc.}
\begin{eqnarray}\label{model}
&v=
(0~0~0~0~0~\frac13~\frac13~\frac23)(0~0~0~0~0~0~0~0)
\nonumber\\
&a_1=
(\frac13~\frac13~\frac13~0~0~\frac13~\frac23~0)
(0~0~0~0~0~0~0~\frac23) \\
&a_3=(0~0~0~0~0~\frac13~0~\frac13)(0~0~0~0~\frac13~
\frac13~\frac13~\frac13).\nonumber
\end{eqnarray}
The resulting gauge group is $SU(3)^3 \times [SO(8)\times
SU(3)]'\times U(1)^4$. The spectrum is shown in Table 1. Note that
this model is not exactly the trinification. Let us break 4
$U(1)$'s completely by VEV's of singlets at a GUT scale.

Note that the three $SU(3)$'s do not have a permutation symmetry
since in the U-sector there appear three more $(\threeb,\bf 3,1)$,
namely only three more lepton humors, and
hence it does not constitute $\Psi_{\rm tri}$
of Eq.~(\ref{trinification}).
The lepton humor itself is not a complete trinification, and hence
with the previous definition of the hypercharge,
$Y=-\frac12(-2I_1+Y_1+Y_2)$, three more lepton doublets survive
down to low energy. To make the theory closer to the SM
spectrum, we must make the extra lepton doublets(of the lepton
humor of the untwisted sector) vectorlike after breaking the
trinification group to the SM gauge group. For this purpose, we
interpret that the 
$SU(3)_h$ from the hidden sector also contributes to
the electroweak hypercharge. QCD is fixed to be
the third $SU(3)_3$. Choosing the second $SU(3)$ as the
weak $SU(3)$, the electroweak hypercharge is defined as,
\begin{equation}\label{Y}
Y=-\frac12(-2I_1+Y_1+Y_2+2I_h-Y_h)
\end{equation}
where $I_i,Y_i\ (i=1,2,h)$ are the generators of $SU(3)_i$ with
eigenvalues $\frac12,-\frac12,0,$ and $\frac13,\frac13,-\frac23$,
respectively, for the fundamental representation {\bf 3}.
Therefore, the SM content of $(\bf 1,\threeb,1)(\bf 1,\threeb)$ of
T8 is the opposite of the lepton humor and can be considered as an
antilepton humor $\bar\Psi_l(T8)$. However, this redefinition of the
electroweak hypercharge changes the electromagnetic content of the
spectrum, and hence changes the bare value of the weak mixing
angle from $\frac38$ to a value not equal to $\frac38$, 
as will be shown later.
Eq.~(\ref{Y}) tolerates $\sin^2\theta_W^0\ne\frac38$ and removes all
the vectorlike representations at a GUT scale, instead of keeping
$\sin^2\theta_W^0=\frac38$ and allow three more lepton doublets at
low energy. Phenomenologically, (\ref{Y}) seems to be better than
the other choice. Of course, a model with six $\Psi_{\rm tri}$'s
and three $\bar\Psi_{\rm tri}$'s keeps both merits, but it will have
more particles than the model presented here. 

The trinification spectrum in T0 has three families. The three
$(\threeb,\bf 3,1)$ in the U-sector is $\Psi_l(U)$ and $(\bf
1,\threeb,1)(\bf 1,\threeb)$ of T8 is a kind of complex conjugate
representation of $\Psi_l$. It is as if we introduce
$\bar\Psi_{l}(T8)$ in the twisted sector T8. Therefore, the neutrino
mass matrix can be made realistic and one can choose a D-flat
direction. For the Higgsino mass matrix, one should consider
$9\times 9$(or $12\times 12$ if we include the vectorlike lepton
pairs from U and T8 also) 
mass matrix from $\Psi_l(U)$, $\bar\Psi_l(T8)$,
and $\Psi_l(T0)$. Suppose that the vectorlike couplings
remove $\bar\Psi_l(T8)$ and one $\Psi_l$, say that of U. 
Then, by giving VEV's to $N_{10}$'s and $N_5$'s, we can break
$SU(3)^3$ to the SM gauge group. Also, we have to break $SU(3)_h$
completely, by giving VEV's to the $Y=0$ and $SU(2)$ singlets in
$(\bf 1,\threeb,1)(\bf 1,\threeb)$ of T8. The vector-like 
representations are expected to be superheavy.(But we want to have
one pair of Higgs doublets light.) The standard
model contents of $SU(3)_c\times SU(2)_W\times U(1)_Y$ nontrivial
representations(using Eq.~(\ref{Y})) from $E_8$ is
\begin{eqnarray}\label{E8}
U&:& \two_{+\frac12},\two_{-\frac12},\two_{-\frac12},\one_{+1}\nonumber\\
T0&:& \two_{+\frac12},\two_{-\frac12},\two_{-\frac12},\one_{+1},
\threeb_{+\frac13},\threeb_{-\frac23},\threeb_{+\frac13},
(\three,\two)_{+\frac16},\three_{-\frac13}\nonumber\\
T1&:& \two_{-\frac16},\one_{+\frac13},\one_{+\frac13},\one_{-\frac23},
\one_{+\frac13},\three_0\nonumber\\
T2&:&
\two_{+\frac16},\one_{-\frac13},\one_{-\frac13},\one_{+\frac23},
\one_{-\frac13},\threeb_0 \\
T6&:&
\two_{-\frac16},\one_{+\frac13},\one_{+\frac13},\one_{-\frac23},
\one_{+\frac13},\three_0\nonumber\\
T7&:&
\two_{+\frac16},\one_{-\frac13},\one_{-\frac13},\one_{+\frac23},
\one_{-\frac13},\threeb_0\nonumber
\end{eqnarray}
where {\bf 3}, {\bf 2}, and {\bf 1} are {\bf (3,1), (1,2)}, and
{\bf (1,1)}, respectively, and the T5 fields are considered below.
The SM contents of $SU(3)_h$ nontrivial representations are
\begin{eqnarray}\label{E8h}
{\rm T3}: {\bf (1,1,1)(1,\threeb)}\oplus{\bf (1,1,1)(1,\three)}&=&
 m^1_{1/3}+ m^2_{-2/3}+ m^3_{1/3}+ \overline{m}^1_{-1/3}+
 \overline{m}^2_{2/3}+\overline{m}^3_{-1/3} \nonumber\\
{\rm T4}: {\bf (1,1,1)(1,\three)}\oplus{\bf (1,1,1)(1,\threeb)}&=&
 \overline{m}^4_{-1/3}+ \overline{m}^5_{2/3}+ \overline{m}^6_{-1/3}
 + m^4_{1/3}+m^5_{-2/3} +m^6_{1/3} \nonumber\\
{\rm T5}: {\bf (\three,1,1)(1,\three)}\oplus{\bf
(1,1,1)(1,\three)}&=&
  n^1_{0}+ n^2_{+1}+ n^3_{0} +\bar n^2_{-1} +n^4_{0} +\bar n^5_{-1}
  +n^6_{0} +n^5_{+1} +n^7_{0}\nonumber\\
{\rm T8}: {\bf (1,\threeb,1)(1,\threeb)}\oplus{\bf
        (1,1,1)(1,\three)}&=& \bar l^1_{+1/2} +\overline{H}^u_{-1/2}
        +\overline{H}^d_{+1/2} +n^8_{0} +e^\prime_{-1} +n^9_{0}
\end{eqnarray}
where we denoted the electroweak hypercharges as subscripts. The
superscripts are just names for particles. $\overline{H}$'s and
$\bar l$ in T8 are doublets. The singlets from T3, T4, and T5 form
vectorlike representations, and we assume that they are heavy.

The chief contributers to the gauge symmetry breaking to
the SM is by the VEV's of $N_{10}$ and $N_5$ in $\Psi_l(U)$ 
and $n_0^8$ and $n_0^9$ of T8. Other $N_{10}$'s and $N_5$'s
can contribute also, but we can assume that they are small.

The bare value of $\sin^2\theta_W=\frac{g^{\prime
2}}{g_2^2+g^{\prime 2}}$ can be calculated at the GUT scale in the
following way. This is useful if everything is embedded
in a simple group, or if all the relevant
representations in consideration are obtainable by tensor
products of some elementary(such as the fundamental representation
in the $SU(5)$ model) representation. 
If $U(1)_Y$ is leaked to other factor
groups, the GUT value of $\sin^2\theta_W$ is not 
the unification value as shown below. 
The definition of $g^\prime$ is $g^\prime Y\equiv
g_1Y_1$ where $Y_1$ is properly normalized. The ratio of $Y$ and
$Y_1$ is $Y=CY_1$, or $g^\prime =C^{-1}g_1$. Therefore, we obtain
$\sin^2\theta_W=\frac{1}{1+C^2}$. On the other hand, the
electromagnetic charge can be similarly normalized,
$eQ_{em}=e_UQ_U$ where $Q_U$ is a universally normalized matrix.
Thus, we have $e^2{\rm Tr.} Q_{em}^2=e_U^2 {\rm Tr.}Q_U^2$.
Similarly, we have $g_2^2{\rm Tr.}(T_3^2)=g_U^2{\rm
Tr.}(T_{U3})^2$. Note that Tr.~$T_{U3}^2=\frac12$ for one doublet
and $\frac{N}{2}$ for $N$ doublets. 
$SU(2)_W$ triplets, quartets, etc. can be properly included. The
unification implies Tr.~$Q_U^2$=Tr.~$T_{U3}^2$. Thus, if the
couplings are unified, i.e. $g_U=e_U$, we obtain
$\sin^2\theta_W=e^2/g_2^2=\frac{{\rm Tr.}~ (T_3)^2}{{\rm Tr.}~
(Q_{em})^2}$. Thus, the bare value of $\sin^2\theta_W$ can be
calculated, using the quantum numbers of
Eqs.~(\ref{E8},\ref{E8h}), to give $\frac27$ if all sets 
are contained in tensor products of some elementary
representations. In fact, ours does not belong to this category,
contrary to the case of trinification, and the bare value of
$\sin^2\theta_W$ is not equal to $\frac27$
but turns out to be $\frac14$.\footnote{There was interest in
$\sin^2\theta_W=\frac14$ before~\cite{onefour}, but that was not
from string theory.}

For the vectorlike $SU(2)$ doublets, we must consider 12 pairs: 3
pairs from U, three pairs from T0, three pairs from T8, and 3
lepton pairs from U and T8. But the essence can be discussed with
just three pairs.

Let us consider the following $3\times 3$ Higgsino mass matrix of
the trinification spectrum,
\dis{ M_{\tilde H}=\left(
\begin{array}{ccc}
  b, &  a, &  a\\
  a, &  b, &  a\\
  a, &  a, & b
\end{array}
\right).\label{mass1} } This form of mass matrix is anticipated if
the spontaneous symmetry breaking of $SU(3)^3$ to the SM gauge
group occurs most symmetrically.

With two Wilson lines, the multiplicity 3 in fact corresponds to
three different fixed points in the third two-torus where no gauge
field is going around. So the three families of $T0$ in fact
corresponds to three different fixed points among 27 fixed points
of $Z_3$ orbifolds. In the orbifold vacua, the well-known
trilinear Yukawa couplings are present for $T_i,T_j,T_k\ (i, j,
k={\rm all\ different})$~\cite{hv87,fiqs}. The trilinear Yukawa
couplings for the fields from the same fixed
points are different from the above couplings of three different 
$T_i$'s. The Higgsino mass matrix arises after assigning
VEV's to $N_{10}$'s, $N_5$'s, $n_0^8$, and $n_0^9$. 
The form (\ref{mass1}) is not a
general one, but it can be a staring point.

\section{Higgsino mass matrix ansatz}

Let us propose the following ansatz for the Higgsino mass matrix,
\begin{equation}\label{ansatz}
Ansatz\ :\ \  {\rm Determinant}\ M_{\tilde H}=0.
\end{equation}

Note that the eigenvalues of (\ref{mass1}) are $b-a,b-a$, and
$b+2a$. There are two solutions of (\ref{ansatz}) with the mass
matrix (\ref{mass1}): one with $b=a$ and the other with $b=-2a$.
The case $b=a$ has two zero eigenvalues and the case $b=-2a$ has
one zero eigenvalue of $M_{\tilde H}$. In view of the mass
hierarchy, one may be attempted to take the solution with $b=a$.

The starting mass matrix (\ref{mass1}) is corrected by the shifts
of VEV's of $n_0^9$'s, $n_0^8$'s, $N_{10}$'s, and $N_5$'s from 
their symmetric points.
Suppose that their shift is not significant. Then, we expect the
mass matrix is given by
 \dis{ M_{\tilde H}=\left(
\begin{array}{ccc}
  b+\lambda_4\epsilon, &  a+\lambda_1\epsilon,
  &  a+\lambda_2\epsilon\\
  a+\lambda_1\epsilon, &  b+\lambda_5\epsilon,
  &  a+\lambda_3\epsilon\\
  a+\lambda_2\epsilon, &  a+\lambda_3\epsilon,
  & b+\lambda_6\epsilon
\end{array}
\right).\label{mass} } where $\lambda$'s are O(1) parameters and
$\epsilon$ is expected to be 
O($\frac{1}{10}\times a$). With the mass matrix
(\ref{mass}), the condition (\ref{ansatz}) gives the eigenvalues,
0, O($\epsilon$), and O($a$) for the masses of Higgsino pairs.
Certainly, this is a solution for the $\mu$ problem, the
doublet-triplet splitting problem and the MSSM problem if all the
$D$ of (\ref{rep2}) and $D^c$ of (\ref{rep3}) are removed at a GUT
scale.

Now, the key theoretical question is how one obtains the ansatz
(\ref{ansatz}). This can be done dynamically by introducing a
scalar field $S$, as the axion chooses the $\bar\theta=0$ vacuum
dynamically. Namely, we make the effective couplings as dynamical
fields, and one of the moduli directions is assumed for this
purpose. For simplicity of the discussion, 
suppose that the moduli $S$
contributes to the $a$(or $b$) couplings but not to the $b$(or
$a$) couplings of Eq.~(\ref{mass}). 
This moduli field settles at a value where
Det.~$M_{\tilde H}=0$.

One obvious reason for Det.~$M_{\tilde H}=0$ is the determinantal
instanton interaction~\cite{thooft}. This determinental
interaction is vanishing if the mass of any matter\footnote{In our
current example the Higgsinos are the matter. For any instanton 
solution absorbing the SM fermions the determinental interaction
does not lead to a condition since the SM fermions are chiral.} 
is zero. For this
to happen, the $SU(3)$ couplings must be strong so that the
instanton interactions are significant. Indeed, for any
supersymmetric $SU(3)$ it is not asymptotically free for $N_F>
3N_c=9$, which is possibly the case if we consider all the
spectrum of the orbifold compactification, as manifested in the
spectrum of Table 1, except for $SU(3)_h$. Above a GUT scale
$M_1$, the theory may not have the asymptotic freedom and becomes
strong at some scale between the $M_1$ and the scale $M_{GUT}$
which can be comparable to the string scale $M_s$. Therefore, the
condition Det.~$M_{\tilde H}=0$ is not unreasonable. If we impose
this condition, one pair of the Higgsinos is massless and survives
down to the the electroweak scale.

But, at a first glance there is no relevant 
$SU(3)$ group for the small instantons
relevant for our scenario. It is because the model is very much
chiral. To have non-chiral pieces, we identify the 9 component
representation of T8 as $\bar\Psi_l$, through the linkage,
$\langle {\bf (3,1,1)(1,3)}\rangle\ne 0$, using the field in T5 .
Then the diagonal subgroup $SU(3)_D$ of $SU(3)_1\times SU(3)_h$ is
unbroken and the 9 component representation of T8 transforms as
${\bf (3,\threeb,1)}$ under $SU(3)_D\times SU(3)_W\times SU(3)_c$
and the $\Psi_l$ of the untwisted sector transforms as ${\bf
(\threeb,3,1)}$ under $SU(3)_D\times SU(3)_W\times SU(3)_c$.
Considering the simplest instanton solution only, the instanton
potential vanishes because the theory is still chiral. To
introduce vectorlike representations absorbed by instantons,
consider a subgroup $SU(2)_D\times SU(2)_W$. This can be an
effective gauge group between the values of 
$\langle N_{10}\rangle$'s and $\langle N_5\rangle$'s.
The values of $\langle N_5\rangle$'s are 
expected to be a factor of $10-100$ smaller than the values of 
$\langle N_{10}\rangle$'s. Consider an
instanton solution between these scales, 
transforming nontrivially under both $SU(2)_D$
and $SU(2)_W$. Then, it absorbs only those fermions transforming
nontrivially under both $SU(2)_D$ and $SU(2)_W$. The matter
connected to this instanton solution must be the Higgsino pairs of
U, anti-Higgsino pairs of T8, and Higgsino pairs of T0. The lepton
and antilepton doublets of U, T8, and T0 do not carry the
$SU(2)_D$ quantum number, and they are not absorbed by this
complex instanton. Then, the determinental interaction of the
remaining 9 pairs of Higgsinos is nontrivial and we can say that
the Higgsino mass ansatz is supported by this instanton calculus.

\section{Doublet-triplet splitting problem}

The form of the mass matrix of the trinification, without
considering the moduli coupling, also applies to the color
triplets $D$ of $\Psi_q$ and $D^c$ of $\Psi_a$
in T0. Let us call these
colored particles as {\it triplets and tripletinos}, meaning color
triplets and their superpartners. But in our model the quarks
carrying $SU(3)_c$ quantum number either does not carry any other
$SU(2)$ quantum number($D,D^c$) or chiral(the SM quark doublets).
So we do not have an argument for a determinental interaction for
the triplitinos. This is the source of the doublet-triplet
splitting in our model. For moduli couplings, the couplings to the
triplitinos must not be aligned to the couplings of Higgsinos.
This kind of different couplings can be possible since the
Higgsino masses arise from $\Psi_l^3$ and $\bar\Psi_l\Psi_l$,
while the triplitino masses arise from $\Psi_l\Psi_q\Psi_a$.
Namely, the gauge group information is different for Higgsino and
triplitino couplings, and it is known that some moduli can
distinguish gauge groups~\cite{ck85}.

There exists the $d=5$ proton decay operator obtained from 
$ql\overline{D}$ and $qqD$, which can be allowable by making
$D$ sufficiently heavy. 
Both of these couplings from the trinification
spectrum can be made absent if we introduce a 
matter parity: $\Psi_{\rm tri}\rightarrow
-\Psi_{\rm tri}$, and others(such as $\Psi_l$ of U) 
= invariant.

\section{Light fermion masses}

In the trinification model, the Yukawa couplings of Higgsinos also
apply to the Yukawa couplings for the quark and lepton families.
However, the quark and lepton masses arise when the light Higgs
boson obtains a VEV. The light Higgs boson is a massless component
obtained from the above mass matrix. This component should be used
for the quark and lepton masses. If we consider the mass matrix
(\ref{mass1}), the massive component is
$H_{u,d}^{GUT}
=\frac{1}{\sqrt{3}}(H_{u,d}^1+H_{u,d}^2+H_{u,d}^3)$ for
$b=a$. The two massless components are orthogonal
to this massive component. Small corrections leave only one
component massless which will be the Higgs fields of
the MSSM. After giving VEV's to these MSSM Higgs fields,
one may also arrive at a flavor democratic mass
matrix. For example, for $Q_{em}=\frac23$ quarks,
\dis{ M_u=\left(
\begin{array}{ccc}
  \frac{m_t}{3}, &  \frac{m_t}{3}, &  \frac{m_t}{3}\\
  \frac{m_t}{3}, &  \frac{m_t}{3}, &  \frac{m_t}{3}\\
  \frac{m_t}{3}, &  \frac{m_t}{3}, &  \frac{m_t}{3}
\end{array}
\right).\label{topmass} } If it is modified little bit by
O($\frac{1}{10}-\frac{1}{100}$) as the Higgsino mass
Eq.~(\ref{mass}) modifies the Higgsino mass Eq.~(\ref{mass1}),
then there is a possibility to have a mass hierarchy for light
fermion families: for example for the up-type quarks, $m_t$, O(a
few times $\frac{1}{100}m_t$), and 0. Here, it is merely a 
speculation, being taken in analogy with the Higgsino mass matrix. 
When supersymmetry is broken
at the TeV scale, the $u$ quark can generate a radiative mass
which can be smaller than the $c$ mass. Therefore, even if Det.
$M_u=0$ at tree level, it is not expected to have a massless up
quark solution for the strong CP problem. A similar argument for
mass hierarchy can be applied to down type quarks and charged
leptons.

For neutrinos, the masses must be much smaller. It can happen
through the seesaw mechanism if the $SU(2)\times U(1)$ singlets
$N_{10}$'s and $N_5$'s obtain large masses near the GUT scale. 
If we did not introduce $\bar\Psi_l(T8)$, 
it is impossible to give large masses to $N_{10}$ and 
$N_5$, because $\Psi_l(T0)^3$ couplings leave
them massless~\cite{pdecay}. 
In our model, there appear $(\bf
1,\threeb,1)(1,\threeb)$ in the T8 sector which allows couplings
rendering $N_{10}$ and $N_5$ superheavy masses. Thus, three
neutrinos can obtain sub-eV masses.

\section{Conclusion}

In conclusion, we proposed an ansatz for the Higgsino mass matrix
toward realizing the MSSM at a high energy scale. This ansatz
can be supported by the short distance dynamics
relevant at the GUT scale such as the small
instanton solutions. We presented this mass matrix
ansatz in a trinification model from a $Z_3$ orbifold 
due to its simplicity in that 
only $SU(3)$'s appear. In principle, this ansatz can be used 
in other unification models. Non-universal moduli couplings 
to the trinification fields are suggested to choose the 
vacuum of our ansatz.

\acknowledgments JEK thanks Hans
Peter Nilles for helpful comments. KSC, KYC, and JEK thank the
Physikalisches Institut of Universit\"at Bonn for the hospitality
extended to them when this work was completed. This
work is supported in part by the KOSEF ABRL Grant to Particle
Theory Research Group of SNU, the BK21 program of Ministry of
Education, and Korea Research Foundation Grant No.
KRF-PBRG-2002-070-C00022.

\begin{table}
\begin{center}
\caption{The massless spectrum of the orbifold
(\ref{model}).}
\begin{tabular}{c|c|c|c}
\hline
sector & twist & multiplicity & fields \\
\hline
U & & 3 & $(\threeb,\three,\one)(\bf 1,1)$ \\
T0 & $V$ & 9 & $(\one,1,1)(\bf 1,1)$ \\
  &  & 3 & $(\threeb,\three, 1)(\one,1)+
(\three,\one,\threeb)(\one,1)+(\bf 1,\threeb,\three)(\one,1)$ \\
T1 & $V+a_1$ & 3 & $(\one,\three,1)(\one,1)+(\three,1,1)
(\one,1)+(\one,1 ,\three)(\one,1)$ \\
T2 & $V-a_1$ & 3 & $(\one,\threeb,1,)(\one,1)+(\threeb,1,1)
(\one,1)+(\one,1,\threeb)(\one,1)$ \\
T3 & $V+a_3$ & 9 & $(\one,1,1)(\one,1)$ \\
  & & 3 & $(\one,1,1)(\one,\threeb)+(\one,1,1)(\one,\three)+
(\one,1,1)(\one,1)+(\one,1,1)(\bf 8,1)$ \\
T4 & $V-a_3$  & 9 & $(\bf 1,1,1)(\bf 1,1)$ \\
  & & 3 & $(\bf 1,1,1)(\bf 1,3)+(\bf 1,1,1)(\bf 1,\threeb)
+(\bf 1,1,1)(\bf 1,1)+(\bf 1,1,1)(\bf 8,1)$ \\
T5 & $V+a_1+a_3$ & 3 & $(\three,1,1)(\bf 1,3)$ \\
T6 & $V+a_1-a_3$ & 3 & $(\bf 1,\three,1)(\bf 1,1)+
(\three,1,1)(\bf 1,1)+(\bf 1,1,\three)(\bf 1,1)$ \\
T7 & $V-a_1+a_3$ & 3 & $(\bf 1,\threeb,1)(\bf 1,1)+(\threeb,1,1)
(\bf 1,1)+(\bf 1,1,\threeb)(\bf 1,1)$ \\
T8 & $V-a_1-a_3$ & 3 & $(\bf 1,\threeb,1)(\bf 1,\threeb)$ \\
\hline
\end{tabular}
\end{center}
\end{table}

\end{document}